\documentstyle[twocolumn,aps]{revtex}

\begin{document}
\draft

\twocolumn[\hsize\textwidth\columnwidth\hsize\csname %
@twocolumnfalse\endcsname

\title{
Heat Transport and the Nature of the Order Parameter in Superconducting
$UPt_3$}

\author{M.R. Norman}
\address{Materials Science Division, Argonne National Laboratory,
Argonne, IL  60439}

\author{P.J. Hirschfeld}
\address{Dept. of Physics, Univ. of Florida, Gainesville, FL 32611}

\maketitle
\begin{abstract}
Recent thermal conductivity data on the
heavy fermion superconductor $UPt_3$ have been interpreted as
offering support for an $E_{2u}$ model of the order parameter
as opposed to an $E_{1g}$ model.  In this paper, we analyze this
issue from a theoretical standpoint including the detailed effects
of Fermi surface and gap anisotropy.  Our conclusion is that although
current data put strong constraints on the gap anisotropy, they
cannot definitively distinguish between these two models.  Measurements on
samples of varying quality could be decisive in this regard, however.
\end{abstract}
]

Well over a decade after the discovery of heavy fermion superconductivity,
the pairing mechanism and even the order parameter symmetry in these
compounds remain controversial.  Early suggestions of pairing
in an unconventional superconducting state, based primarily on analysis
of transverse ultrasound measurements in
$UPt_3$,\cite{shivaram} were bolstered
more recently
by the discovery of a complex phase diagram for this system in
applied magnetic field and pressure.\cite{review}
(Here we take ``unconventional''
to imply the existence of additional broken symmetries beyond the
usual gauge U(1) broken in classic superconductors.\cite{sigrist})

Several current Ginzburg-Landau (GL) theories of the $UPt_3$
phase diagram attribute the existence of multiple superconducting
phases to two nearly degenerate superconducting states, either
(i) split by a symmetry-breaking field, such as the ordered antiferromagnetic
moment in the basal plane, or (ii) ``accidentally''
degenerate.\cite{review}  Such
theories can at the same time describe qualitatively  the anisotropy
of the superconducting state, insofar as the GL parameters can be
chosen to stabilize an order parameter at low temperatures and fields
which allows for a larger number of quasiparticle excitations with
wave vector in the basal plane.  Such a state is strongly indicated
by analyses\cite{ph1,srink} of both ultrasound\cite{shivaram}
and thermal conductivity measurements.\cite{flouquet,tai1}
Beyond this crude statement, little is known for certain about the
exact anisotropy or even the symmetry of the superconducting state of $UPt_3$.

Recently, Lussier et al\cite{tai1} have argued that thermal conductivity
measurements
can shed further light on these questions.  They showed that
the electronic heat current dominates the phononic current
down to low temperatures for their high quality samples,
and that the relaxation rate  $1/\tau_k$ in
the normal state is nearly isotropic.
Furthermore, their measurements imply the existence of large anisotropy
in the superconducting state which does not simply reflect
normal state anisotropy;
together with transverse ultrasound measurements, these data
provide convincing evidence for a highly anisotropic gap in $UPt_3$.

In order to determine the actual gap anistropy for $UPt_3$,
it is necessary to go further and attempt to model the data.
While it has been stated that such fits to transport properties cannot
be expected
to fix the detailed anisotropy due to uncertainties in the form
of the impurity scattering amplitude,
Fledderjohann and Hirschfeld\cite{FH} argued recently that ratios of
transport coefficients should lead to more robust conclusions since
they can depend only weakly on the relaxation times.  They therefore
focussed on
the ratio $\kappa_c /\kappa_b$ between the conductivities
measured for heat currents directed along the c-- and b--axes,
respectively, comparing
the data of Lussier et al\cite{tai1} to weak-coupling BCS calculations
using order parameters representative of the $E_{1g}$ and $E_{2u}$
symmetry classes of the $D_{6h}$ space group of the hexagonal crystal.
While both states have lines of order parameter nodes
(and hence higher density of
excited quasiparticles) in the basal plane, the $E_{2u}$ state has
point nodes along the c-axis where the order parameter vanishes quadratically,
in contrast to the linear behavior in the $E_{1g}$ state.
In consequence, the thermal
conductivity (and indeed all current-current correlation functions)
was found to be
isotropic in the $E_{2u}$ state over a spherical Fermi surface,
despite the intrinsic anisotropy of the superconducting state.
Ellipsoidal Fermi surfaces do not change the value of the normalized
conductivity ratio $(\kappa_c/\kappa_{Nc}) /(\kappa_b/\kappa_{Nb})$
from unity in this state ($N$ refers to the normal state),\cite{ell}
but it is clear that the true hexagonal crystal
structure will do so.
Furthermore,  is not clear whether this result is
specific to the particular $E_{2u}$ state analyzed, or would hold for a
more general $E_{2u}$ state.  Understanding the extent to which these
factors might improve the agreement with the large measured anisotropy
is crucial to the $E_{2u}$ scenario proposed by Sauls and
Norman\cite{review} in
which several problems characteristic of GL theories of type (i) above
are resolved.

In this paper, we study the influence of both Fermi surface anisotropy and
gap anisotropy on
superconducting state transport coefficients, focussing on the thermal
conductivity data of Lussier et al.\cite{tai1,tai2}  In the first part, we
use a simple ellipsoidal Fermi surface fit to normal state transport data and
analyze all gap functions represented by ellipsoidal harmonics up through L=5,
treating various impurity scattering rates, impurity phase
shifts, and inelastic scattering effects.  First, we find that a finite,
non-zero T=0 value of the thermal conductivity ratio,
$\kappa_c(0)/\kappa_b(0)$,
of intrinsic origin occurs for a number of harmonics, not just those
of $E_{2u}$ symmetry.  Second, we find that the data can be fit reasonably
well by gaps of both $E_{1g}$ and $E_{2u}$ symmetry, with the latter fitting
slightly better than the former, although in neither case is a pure harmonic
realized.
These fits could be differentiated more clearly by (1) extending
the measurements to lower temperatures or (2) by increasing or decreasing
the impurity scattering rate, that is, by analyzing cleaner or dirtier samples.
Although fits using ellipsoidal harmonics may be somewhat unrealistic, they
allow us to obtain some useful analytical results, and determine
the qualitative features of order parameter anisotropy with some
confidence.

In the second part, we turn to the more general case, using the multi-sheeted
Fermi surface predicted from local density approximation (LDA)
calculations\cite{nabc} which is in reasonable agreement with deHaas-vanAlphen
(dHvA) experiments,\cite{dhva} up to an overall mass renormalization.
Two types of gap functions are analyzed:  Fermi surface harmonic\cite{allen}
and tight binding\cite{putikka,konno}.  In neither case is an adequate fit
found
to the data for either $E_{2u}$ or $E_{1g}$ with single basis functions,
although one of the tight
binding gap functions of $E_{1g}$ symmetry has some promise.  In the Fermi
surface harmonic case, this poor agreement is due to the large number of nodes
these functions
possess which is unlikely to arise out of any microscopic gap equation.  In the
tight binding case, this is likely due to the use of a single basis function.
Use of a mixed basis set in the tight binding case leads to a good
correspondence to the data in the $E_{1g}$ case.  So far, we have not found a
comparably good fit for the $E_{2u}$ case.

\section{Order Parameters and Fermi Surfaces}

Although a variety of models have been proposed for the order parameter of
$UPt_3$,
we concentrate here on the most popular model, that of a two-dimensional
group representation.  The two variants most commonly explored have been
the $E_{1g}$ model\cite{joynt} and the $E_{2u}$ model\cite{review}.  For a
spherical Fermi surface, the gap function can be represented by spherical
harmonics.
A function of $E_{1g}$ symmetry first occurs in the L=2, M=1
representation (d-wave).
The $E_{2u}$
case is more subtle since it is an odd parity gap and therefore a pseudo-spin
triplet.\cite{pseudo}
The proposed $E_{2u}$ model  assumes that
the gap is a pure spin triplet with only one component ($S_z=0$)
condensed, however.  In this case, $E_{2u}$ first occurs for $Y_{32}$ (f-wave).
The $E_{2u}$ model based on $Y_{32}$ was originally proposed\cite{norman} since
(i) its nodal structure was similar to the previously considered $E_{1g}$ model
based on $Y_{21}$, with line nodes perpendicular to the c-axis and point nodes
along the c-axis as indicated by transverse ultrasound\cite{shivaram} as well
as point contact spectroscopy,\cite{point} and (ii) it has an upper critical
field anisotropy consistent with experimental data\cite{shiv2} given the
$S_z=0$ orientation of the triplet order parameter, as demonstrated earlier by
Choi and Sauls\cite{choi} (singlet order parameters give an incorrect
anisotropy).  Sauls\cite{review} in turn showed that this model
solved a major problem of the previously considered $E_{1g}$ model, in that it
could explain the existence of a tetracritical point in the H-T phase diagram
for all orientations of
the magnetic field as observed experimentally, at least for axial symmetry.
Recently, Park and Joynt\cite{park} have proposed that $E_{1g}$
can avoid the problem of an incorrect upper critical field anisotropy if
the normal state Pauli susceptibility has opposite anisotropy to the observed
normal state susceptibility (the latter likely being van Vleck dominated).  It
can also give a phase diagram which has a near tetracritical point for certain
choices of the GL coefficients, with the additional claim that it gives a
better explanation of the pressure-temperature phase diagram than $E_{2u}$.

A potential method of resolving these controversies would be to obtain more
knowledge of the actual form of the gap anisotropy.  The $E_{1g}$ model
has a linear dispersion of the quasiparticle energies about the point nodes,
whereas the $E_{2u}$ model has a quadratic dispersion.  This can have a
significant effect on transport quantities, as pointed out by Yin and
Maki.\cite{yin}  Fledderjohann and Hirschfeld\cite{FH} exploited this to
show that the thermal conductivity anisotropy ratio, $\kappa_c(0)/\kappa_b(0)$,
is small for the $E_{1g}$ case (it would be zero in the clean limit),
but is unity for the $E_{2u}$ case, at least for an
ellipsoidal Fermi surface, with the data of Lussier et al\cite{tai1} lying
between these two results but being more consistent with $E_{1g}$ than
$E_{2u}$.
This in turn motivated Lussier et al to take data at lower
temperatures, where they conclude that the extrapolated T=0 anisotropy ratio
of about 0.5 is probably intrinsic and thus consistent with an $E_{2u}$
model.\cite{tai2}

The above analysis of Fledderjohann and Hirschfeld\cite{FH} was based on a
particular spherical harmonic form of the order parameter on an ellipsoidal
Fermi surface.  For a real metal like $UPt_3$, we would expect that the actual
order parameter is more complicated, just as we know that the actual Fermi
surface is multi-sheeted and shows strong deviations from axial
symmetry.\cite{dhva}  The latter is particularly important since the $E_{2u}$
result $\kappa_c/\kappa_b=1$ is a consequence of axial symmetry.

To analyze this in more detail, we first consider the simple ellipsoidal
case treated previously, but look at other harmonics besides $Y_{21}$ and
$Y_{32}$.  The conversion from spherical to ellipsoidal harmonics can be
achieved by replacing $\sin(\theta)$ by $\sin(\theta)/\sqrt{m_r}$
and $r^2$ (previously unity) by $\cos^2(\theta)+\sin^2(\theta)/m_r$\cite{fsh}
where the mass ratio, $m_r=m_{\perp}/m_c$, is equal to 2.8 based on normal
state
transport data.\cite{tai1}  Note that this conversion simply multiplies
$Y_{21}$ and $Y_{32}$ by an overall constant, so the results for these two
cases are independent of the mass ratio.  This is not true in general.
For the $E_{1g}$ case, the next higher harmonic to consider is $Y_{41}$;
for the $E_{2u}$ case ($S_z=0$), $Y_{52}$.  Although such
higher harmonics seem exotic, they do play a significant role in certain
microscopic theories.\cite{hund}

For the real Fermi surface case, we utilize the surface obtained from an LDA
calculation.\cite{nabc,nlda}  This surface, which is
shown in symmetry planes of the zone in Fig. 1, is in reasonable
agreement with dHvA data, except for mass renormalization
effects.  The mass renormalization would play no role here unless
it was anisotropic.  Unfortunately, there is not enough data available to
model this anisotropy, although the lack of observation of dHvA orbits for
fields along the c axis suggests that the renormalization is anisotropic and
will act to increase the mass ratio, $m_r$.\cite{dhva}  To check this,
we calculated
$<v_c^2>/<v_b^2>$ where $v_b$ and $v_c$ are the Fermi velocities and $<>$
is an average over the Fermi surface determined by using a linear tetrahedron
decomposition of the Brillouin zone.\cite{tet1,tet2}  This quantity, equal to
the ratio $\kappa_{Nc}/\kappa_{Nb}$ given the observation of
isotropic relaxation time,\cite{tai1} is 2.8 from thermal conductivity
and 2.7 from resistivity.\cite{tai1}  The LDA calculation gives 2.1.  This is
consistent with the above observation that the mass renormalization anisotropy
acts to increase $m_r$.  Since we are comparing in this paper to data
normalized to the normal state value, we ignore this mass ratio discrepancy
in this paper, although we caution that it could influence some of the
results presented here.  In particular, an alteration in Fermi surface topology
would certainly change the gap anisotropy, but even a momentum
dependence of the mass renormalization would affect the results since this
would act to alter the weighting of various momentum vectors in the equations.

The problem of what gap function to use for the real Fermi surface case is
a more complicated issue.  Allen showed a number of years ago that functions
could be constructed on the Fermi surface which were orthornormal and so
could be used as basis functions, which he labeled Fermi surface
harmonics.\cite{allen}  They can be obtained from the spherical harmonics
by replacing $k_i$ by $v_i$,\cite{fsh} although there are more Fermi surface
harmonics for a given power of $v_i$ than there are spherical harmonics since
the lattice symmetry is discrete rather than continuous.  A complication
is that the relative size of the gap function on independent sheets of the
Fermi surface cannot be determined outside the context of a microscopic
theory of the superconductivity.  In fact, the problem is a more general one
since the simple gap functions treated here are likely to be modulated at each
k vector by some complicated function (of $A_{1g}$ symmetry) that contains
the affects of wavefunction anisotropy, etc.
We further note that the Fermi surface harmonics will
be small when the Fermi velocities are small.  On the other hand, we would
anticipate that in most microscopic theories, the gap function will be
large where the f electron weight is highest, which is also where the Fermi
velocities will in general be lowest.  Keeping these caveats in mind, we define
Fermi surface harmonics for the $E_{1g}$ case and $E_{2u}$ case by replacing
$k_i$ by $v_i$ in the $Y_{21}$ and $Y_{32}$ spherical harmonics.  That is,
the modulus of the $E_{1g}$ gap is $v_cv_r$ and the $E_{2u}$ gap $v_cv_r^2$
where $v_r^2=v_a^2+v_b^2$.\cite{fsh2}  The nodal
structure of these functions are very complicated given the complicated Fermi
surface geometry.  There
are many points on the Fermi surface where the velocity vector points either
along or perpendicular to the c-axis.  In any of these cases, the $E_{1g}$
and $E_{2u}$ Fermi surface harmonics will vanish.

An alternate set of basis functions can be generated by tight binding
expansion.  In the square lattice case, the lattice vectors of type (1,0)
lead to a d-wave state of the form $\cos(k_x)-\cos(k_y)$, which is currently
the leading model being explored for high temperature cuprates.  For the
hexagonal closed packed case, $E_{1g}$ and $E_{2u}$ first appear for
primitive lattice vectors
of the type (0,1,1).  These can be generated from the next near neighbor
basis functions listed by Putikka and Joynt\cite{putikka} by multiplying
their $E_{1u}$ and $E_{2g}$ functions by $\sin(k_zc)$.  They are for $E_{1g}$
\FL
\begin{eqnarray}
f_1 = \sqrt{2}\sin(k_zc)\cos(\frac{1}{2}k_ya)\sin(\frac{\sqrt{3}}{2}
k_xa) \nonumber \\
f_2 = \frac{2}{\sqrt{6}}\sin(k_zc)(\sin(k_ya)+\sin(\frac{1}{2}k_ya)
\cos(\frac{\sqrt{3}}{2}k_xa))
\end{eqnarray}
and for $E_{2u}$
\FL
\begin{eqnarray}
f_1 = \frac{2}{\sqrt{6}}\sin(k_zc)(\cos(k_ya)-\cos(\frac{1}{2}k_ya)
\cos(\frac{\sqrt{3}}{2}k_xa)) \nonumber \\
f_2 = \sqrt{2}\sin(k_zc)\sin(\frac{1}{2}k_ya)\sin(\frac{\sqrt{3}}{2}k_xa)
\end{eqnarray}
with a gap modulus of $\sqrt{f_1^2+f_2^2}$ for an assumed gap of
the form $f_1+if_2$ (the 1,i state).
Both functions
have line nodes in the $k_z=0$ and $k_z=\pi/c$ planes.  In addition, the
$E_{1g}$ function has point nodes with linear dispersion along all three
symmetry axes ($\Gamma-A$, $M-L$, $K-H$), whereas $E_{2u}$ has quadratic
point nodes along $\Gamma-A$ and linear point nodes along $K-H$.
We note that the
basis functions listed by Putikka and Joynt for the near neighbor case
are not properly invariant under reciprocal lattice translations due to
the non-symmorphic nature of the $UPt_3$ lattice (that is, the near neighbors
are separated by a non-primitive translation vector).  This problem has been
addressed by Konno and Ueda.\cite{konno}  Under the highly simplistic
assumption that the phase of the single particle wavefunctions
on the two sites in the unit cell is determined by a simple near neighbor
interaction, they were able to generate analytic near neighbor basis functions
which have
proper translational symmetry.  In this paper, we use their $\Gamma_6^+$
($E_{1g}$) and $\Gamma_5^-$ ($E_{2u}$) basis functions.  These, in fact,
can be generated from the tight binding basis functions discussed above by
replacing $\sin(k_zc)$ by $\sin(k_zc/2)/|\phi|$ where $\phi$ is the Fourier
transform of the near neighbor distance vectors projected onto the basal
plane
\begin{equation}
\phi = \frac{1}{\sqrt{3}}(e^{i\frac{k_xa}{\sqrt{3}}}+2\cos(\frac{k_ya}{2})
e^{-i\frac{k_xa}{2\sqrt{3}}})
\end{equation}
($\phi$ is complex since the lattice is non-symmorphic).  The effect
of this is to remove the line nodes in the $k_z=\pi/c$ plane and the linear
point nodes along $K-H$.

\section{Thermal conductivity}
The thermal conductivity $\kappa$  in the presence of impurities
is evaluated using a Kubo formula
for the heat-current correlation function as in the original treatment
for an s-wave superconductor by Ambegaokar and Tewordt.\cite{amb}
This treatment was generalized to unconventional
states  by several
groups, giving results for a spherical Fermi surface
and model p- and d-wave states which agreed qualitatively
with experiment.\cite{ph1,srink,Monien,ph2}
In the limit of vanishing impurity concentrations, identical results
were also obtained by Arfi et al\cite{Arfi}
using  a transport equation method.

The Kubo formula approach begins with an impurity-averaged  single-particle
matrix propagator

\begin{eqnarray}
\underline{g}({\bf k},\omega)
 =
\frac{\tilde{\omega} \underline{\tau}
^0 + \xi_{\bf k}
\underline{\tau}^3 + \underline{\Delta}_{\bf k} }{\tilde{\omega}^2 -
\xi_{\bf k}^2 -
|\Delta_{\bf k}|^2}
\end{eqnarray}
\noindent
where $\underline{\tau}^i$ represent the Pauli matrices spanning
particle-hole space. Here, we have already
exploited the approximate particle-hole symmetry of the
normal state, as well as
the symmetries  of the gap functions which lead to  vanishing off-diagonal
scattering self-energy contributions.
In this limit, only self-energy contibutions
to the frequency $\omega$, namely
$\tilde{\omega}= \omega - \Sigma_0$ need
to
be included.\cite{ph2}
The self-energy $\Sigma_0$ due to the elastic impurity scattering is
treated in a self-consistent $T$-matrix approximation and
is given by
$\Sigma_0
 =
\Gamma G_0/(c^2-G_0^2),$
\noindent
where $\Gamma=n_i n/(\pi N_0)$ is the unitarity
limit scattering rate  depending on the concentration of defects $n_i$,
the electron density $n$, and the density of states at the Fermi level $N_0$.
The quantity $c\equiv \cot \delta_0$
parameterizes the scattering strength of an individual
impurity through the s-wave phase shift $\delta_0$.  In this work we
consider primarily unitarity limit scattering $c=0$ since it is clear
that weak scattering will lead to a weak temperature dependence inconsistent
with experiment for the states in question.\cite{coffey,pethick}
The integrated propagator is
$
G_0
 =
(1/2\pi N_0) \sum_{\bf k}  \mbox{Tr} \{\underline{
\tau}^0\underline{g}({\bf k},\omega)\}.
$
\noindent
The equation for the self-energies are now solved self-consistently.  We ignore
the complication of resolving the gap equation in this paper since all results
are scaled to $T_c$ and impurity corrections are small.
The gap itself is given by
\begin{equation}
\Delta_{\vec k}(T) = \Delta_{BCS}(T) e^{-<f_{\vec k}^2\ln|f_{\vec k}|>
/<f_{\vec k}^2>} f_{\vec k}
\end{equation}
where $\Delta_{BCS}(T)$ is the BCS gap and $f_{\vec k}$ is one of the basis
functions (or mixtures thereof) described in the previous section.

The bare heat current response is given by a convolution of
the Green's function $g$ with itself at zero external frequency and wave vector
weighted with the bare heat current vertex
$\omega {\bf v}_{{\bf k}} \underline{\tau}^3$.\cite{amb}
Impurity scattering vertex corrections to current-current correlation functions
have been shown to vanish identically for even
parity states ($\Delta_{\bf k} = \Delta_{-{\bf k}}$).\cite{ph2} Even for
odd parity states,   such corrections vanish
in the unitarity limit.
For the diagonal thermal conductivity tensor one obtains
\begin{eqnarray}
\frac{\kappa_i(T)/T}{\kappa_{N,i}(T_c)/T_c}
=
\frac{3\Gamma_N}{4\pi^2}
\int_0^{\infty}\frac{d\omega}{T}\left(\frac{\omega}{T}
\right)^2 {\rm sech}^2\left( \frac{\omega}{2T}\right)
K_i(\omega , T)\\[10pt]
\label{K}
K_i(\omega , T)
 =
\frac{1}{\tilde{\omega}^\prime
\tilde{\omega}^{\prime\prime}}
{\rm Re} \int {dS_{\bf k} \over |{\bf v_k}|}
\frac{{\bf v_k}_i^2}{<{\bf v_k}_i^2>}
\cdot \frac{\tilde{
\omega}^2 + |\tilde{\omega}|^2 - 2|\Delta_{\bf k}|^2}{\sqrt{\tilde{\omega}^2
- |\Delta_{\bf k}|^2}}
\end{eqnarray}
\noindent
where ${\tilde\omega}^\prime$ and ${\tilde\omega}^{\prime\prime}$ are the
real and imaginary parts of $\tilde \omega$ and $\Gamma_N\equiv\Gamma/(1+c^2)$
is the normal state scattering rate.
Here $dS_{\bf k}$ is the area measure on the Fermi
surface, and $v_{\bf k}$ is the Fermi velocity.

For a complete description of the data, we must take into account the
effects of inelastic scattering.  This is known to vary as $b T^2$ times the
elastic rate in the normal state, with $b \simeq 4/K^2$.\cite{tai1}
This effect can be included in the above equations by replacing $\Gamma$ by
$\Gamma(1+bT^2)$.
In the superconducting state,
we can make the ansatz that the inelastic rate varies as $b T^3/T_c$ since
the number of quasiparticles varies as $T/T_c$ at low temperatures due to the
line nodes in the gap.\cite{quinlan}  The
exact form of this makes little difference, since
by far the largest effect inelastic scattering has is on $\kappa_N(T)$.
Therefore, for practical purposes when comparing to data normalized to its
value at $T_c$, one
can simply scale the result of Eq. 6 by $(1+bT_c^2)/(1+bT^2)$.

Finally, we note that $UPt_3$ has a split superconducting phase
transition.  In the $E$ models considered here, this splitting is assumed to
be due to the weak antiferromagnetism which has orthorhombic symmetry.  Its
effect is to cause only one of the two $E$ components to condense at the
upper phase transition.  Thus, in the region between the upper ($T_{c+}=0.50
K$)
and lower ($T_{c-}$=0.44 K) transitions, the point nodes along the c axis
become
line nodes perpendicular to the basal plane.  This explains the lack of gap
anisotropy in the thermal conductivity observed in this region.  Below
$T_{c-}$, the second $E$ component condenses, and the anisotropy begins to
occur, as observed.  Modeling this is complicated since the calculation would
have to be performed for orthorhombic symmetry with two different gaps and
appropriate domain averaging performed.  We therefore take the approach of
previous work which ignores this symmetry breaking but normalizes
$\kappa_c/\kappa_b$ to its value at $T_{c-}$.\cite{tai1,FH,tai2}  This
normalization does not work so well, though, when comparing to the
individual $\kappa_i(T)/T$ themselves, since the thermal conductivity
does change above $T_{c-}$.  We have found that normalizing $\kappa_i(T)/T$
to its value at $T_{c0}=0.47 K$, the average of $T_{c+}$ and $T_{c-}$ (the
``hexagonal'' $T_c$), works quite well in this regard.  Obviously, one cannot
take too seriously the results in the immediate vicinity of $T_c$ until
the effects of the symmetry breaking field are properly included.

\section{Results}

\subsection{Ellipsoidal harmonics}

Results for $\kappa_c/\kappa_b$ are shown in Fig. 2 for all harmonics
through L=5 on an ellipsoidal Fermi surface with $m_r=2.8$, compared to the
experimental results of Lussier et al.\cite{tai2}  This quantity is normalized
to its value at $T_{c-}$ as discussed above.  The results were generated
with an impurity scattering rate in the unitarity limit of 0.1 $T_c$,
consistent with experimental data (particularly with the observation of a
residual linear specific heat coefficient of 0.16 the normal state
value).\cite{tai2}  To understand
these results more clearly,
we have analytically calculated
\begin{equation}{\kappa_c(0)\over\kappa_b(0)}=
\lim_{\omega\rightarrow0}
{
{{\rm Re}\langle v_{c}^2 \sqrt{\omega^2-\Delta_k^2} \rangle}
\over
{{\rm Re}\langle v_{b}^2 \sqrt{\omega^2-\Delta_k^2} \rangle}
}
\end{equation}
\noindent
in the
clean limit
for a spherical Fermi surface,
and show these results in Table 1.
We see that
harmonics of the form $Y_{LL}$, which have only point nodes along the c axis,
give a divergent ratio.  On the other hand, only two of the remaining
harmonics,
$Y_{10}$ and $Y_{21}$, give a ratio of zero (the non-zero value in Fig. 2a is
due to impurity-induced gaplessness\cite{FH}), with the rest giving an
intrinsic
non-zero ratio.  In particular, we note that harmonics of the form $Y_{L0}$
only have line nodes, so a finite, non-zero ratio is not something just
associated with quadratic point nodes or with gaps of $E_{2u}$ symmetry.
We also note that no pure harmonic provides a good fit to the observed
anisotropy,\cite{fsh} although some higher harmonics give adequate fits.

To study this further, we have looked into the possibility of mixed solutions.
For $E_{1g}$, we included mixing of $Y_{21}$ with $Y_{41}$; for $E_{2u}$,
$Y_{32}$ with $Y_{52}$.\cite{y54}
Typical results for $\kappa_c/\kappa_b$ are shown in Fig. 3a, with
the coefficients roughly optimized to fit the data.  Both fits give
a reasonable description of the data, with the lowest temperature data
intermediate between the two results (it should be remarked that the
error bars on the experimental $\kappa_c/\kappa_b$ are about 15\% at
low temperature\cite{tai2}).  The $E_{1g}$ fit can be greatly improved at
lower temperature
by going to a larger scattering rate of 0.3 $T_c$, but the individual
$\kappa_i/T$ in this case are in poor agreement with experiment.
Altering the scattering
phase shift from the unitarity value of $\cot(\delta_0)$ of 0 to 0.2
slightly improves things at the lowest temperatures, but this is probably
not significant given the experimental error bars.  For the $E_{2u}$ case,
lowering the scattering rate by a factor of ten only slightly suppresses the
ratio and only for temperatures below where experimental data exist (the
same slight suppression also occurs by increasing $\cot(\delta_0)$).  In
Fig. 3b, we compare these fits to the individual $\kappa_i/T$, normalized to
their value at $T_{c0}$ as discussed previously.
As can be seen,
both $E_{1g}$ and $E_{2u}$ provide good fits to the data, with
the $E_{2u}$ fit being slightly superior.  An interesting point is that the
experimental $\kappa_i/T$ are linear in temperature down to the lowest measured
temperatures.  This behavior cannot continue indefinitely since $\kappa_i/T$
would be zero at a temperature larger than zero (that is, $\kappa_i/T$
must flatten off).  The calculated curves, though, predict that this
flattening should occur in the measured temperature range in contradiction
with experiment.  As suggested by
the authors of Ref. \onlinecite{tai2}, the calculated low temperature
behavior can be improved by reducing $\Gamma$ to roughly  one tenth
its normal state value.  Although this does improve
the fit at low temperatures,
it leads to a substantial deviation from the data at higher temperatures.
In fact, we have found that the value $\Gamma/T_c \sim 0.1$ (the normal state
value) gives roughly the
best fit over the entire temperature range below $T_c$.  We have also found
that altering the scattering phase shift from the unitarity value of
$\pi/2$
does not improve the fit in this regard, at least for small values of
$\cot(\delta_0)$.  If this discrepancy between the low temperature and high
temperature behavior
is taken at face value, a strong
temperature dependence of either $c=\cot(\delta_0)$ or $\Gamma$
must be assumed.  Although dynamical scattering effects could easily
influence the phase shifts in this way, we have been unable to
find a satisfactory phenomenological explanation of the data
in these terms.  While a T-dependent parametrization of $\Gamma$
could possibly account for the discrepancy, we have no physical
understanding of how such effects could arise.

The gap functions in the $E_{1g}$ and $E_{2u}$ cases are plotted as a
function of
polar angle in Fig. 4.  Both gap functions look similar (except for
the different dispersions around the point nodes at zero degrees).  This
indicates that the primary determinant of the thermal conductivity is the
overall shape of the gap function.  We also note that the maximum gap occurs at
a polar angle of 52 degrees for $E_{1g}$ and 49 degrees for $E_{2u}$.
Not only are these angles close, but they are also close to the
angle of 54 degrees that the vector connecting near neighbor uranium atoms
makes with the c axis.  This could be taken as indirect evidence
that the electrons in the Cooper pairs reside at near neighbor sites as
would be predicted by microscopic models based on antiferromagnetic spin
fluctuations.

We conclude this part by remarking that both the $E_{1g}$ and $E_{2u}$ models
can explain the data.  One way to more clearly distinguish between the two
would be to carry the experiments to lower temperatures, although
given experimental error bars, it may be difficult to conclude anything
definitive.  At the least, one would hope to see $\kappa_i/T$ flatten off.
Perhaps a better way would be to degrade the quality of
the sample.  In the presence of a finite concentration of impurities,
states with line nodes yield a linear term in the thermal conductivity,
$\kappa_i(T)\sim  T/\Delta_0$ at the lowest temperatures. For generic
configurations of the thermal current ${\bf j}_Q$ and the line nodes,
the proportionality constant is actually independent of the
impurity scattering rate to leading order, yielding a universal
value for the low-T thermal conductivity analogous to the electrical
conductivity result found by P.A. Lee.\cite{PALee}  In the final
stages of writing, we received a paper from Graf et al.\cite{saulskappa}
in which
this result was obtained independently and explored in some detail.
\vskip .2cm
Because of impurity-induced gapless effects of this type,
the anisotropy ratio $\kappa_c(0)/\kappa_b(0)$ is always finite even for
states like $E_{1g}$, as mentioned above.  In such a situation, an estimate
of the anisotropy may be performed by considering Eq. 7 in the gapless
regime, i.e. take $\tilde\omega=a\omega+i\gamma$, with $a$, $\gamma$
constant.  We then find
$\kappa_c(0)/\kappa_b(0)=\langle v_c^2 F_k^3\rangle/
\langle v_b^2 F_k^3\rangle$, with $F_k=(\gamma^2+\Delta_k^2)^{-1/2}$.  For
the $Y_{21}$ case ($E_{1g}$) in spherical symmetry, we find
$\kappa_b(T)\sim T/\Delta_0$, $\kappa_c(T)\sim \gamma T/\Delta_0^2$, giving
$\kappa_c(0)/\kappa_b(0)\simeq 2\gamma/\Delta_0$ for small concentrations.
The residual broadening $\gamma$ is found by solving the
transcendental equation $c^2+\gamma^2\langle F_k\rangle^2=
\Gamma\langle F_k\rangle$, and yields a square root dependence on
concentration, $\kappa_c(0)/\kappa_b(0)\sim(\Gamma/\Delta_0)^{1/2}$
up to logarithmic corrections in the unitarity limit $c=0$.  In
Fig. 5, we plot the impurity
concentration dependence of the anisotropy ratio in both $E_{1g}$
and $E_{2u}$ cases for $T$=0.02 and 0.2 $T_c$.  We note the qualitatively
stronger dependence of the anisotropy ratio on impurity scattering rate
for the $E_{1g}$ case as
compared to the $E_{2u}$ case at low temperatures.

\subsection{Fermi surface harmonics}

The above analysis assumes an ellipsoidal Fermi surface.  The actual Fermi
surface for $UPt_3$ is very complicated and could substantially alter the
above conclusions.  In this
part, we present results using Fermi surface harmonic gap functions as
described in the first section with the Fermi surface shown in Fig. 1.  We
again
assume an impurity scattering rate of 0.1 $T_c$.  Results for
$\kappa_c/\kappa_b$ are shown in Fig. 6a and for $\kappa_i/T$ in Fig. 6b.
The $E_{1g}$ case gives a fair representation of the anisotropy ratio,
especially at higher temperatures.  Both cases, though, predict too
large $\kappa_i/T$ at lower temperatures.  This occurs
since the Fermi surface harmonic gap functions have a very large number
of nodes.  These nodes are due to the complicated Fermi surface, which
has many points on it where the velocity vector is either parallel or
perpendicular to the c axis, either case in which the $E_{1g}$ and $E_{2u}$
Fermi surface harmonics vanish.  This large number of nodes is unlikely
to arise out of any microscopic gap equation since such a solution would
not have an optimal condensation energy.  This indicates that Fermi surface
harmonics are unlikely to be useful in modeling heavy fermion
superconductors.  Because of this, and since the addition of higher order
harmonics will not reduce the number of nodes, we have not explored
using mixed Fermi
surface harmonics as done in the previous subsection for ellipsoidal
harmonics.

\subsection{Tight binding functions}

We next present results for tight binding gap functions.  These functions
represent short range interactions in the lattice, and therefore are
more likely to arise out of a microscopic gap equation than the Fermi
surface harmonics.  Results for the functions based on (0,1,1) type
primitive lattice vectors are shown for $\kappa_c/\kappa_b$ in Fig. 7a
and $\kappa_i/T$ in Fig. 7b.  Although the magnitude of $\kappa_i/T$ is
improved over the Fermi surface harmonic case, the anisotropy ratio is in poor
agreement with experiment.  We therefore turn to results using the Konno-Ueda
functions based on near neighbor interactions\cite{konno} (Figs. 8a and 8b).
The observed anisotropy ratio is intermediate between the $E_{1g}$
and $E_{2u}$ cases.  Although $\kappa_i/T$ is too large at lower
temperatures for $E_{2u}$, it is not too bad for $E_{1g}$.  Perfect
agreement with experiment would not be expected anyway since the above
basis functions do not take into account the complicated single particle
wavefunctions which occur in heavy fermions due to f orbital degeneracy
and f-ligand hybridization.
It is interesting to note that the near neighbor tight binding
functions provide the best overall comparison to the data for those
functions we have analyzed on the real
Fermi surface, since this reinforces the idea of near neighbor pairing
that was suggested from the ellipsoidal results above.
Given this, it will interesting in the future to calculate $\kappa$ for
recent microscopic
models which do not involve such pairing.\cite{hundx}

As in the ellipsoidal harmonic case,
to test the idea of whether a mixture of tight binding basis functions would
improve
the results, we have done calculations mixing the two tight binding functions
considered above for each symmetry.  Only a rough optimization of the mixing
coefficients could be determined due to calculational demands.  These results
are presented in Figs. 9a and 9b.  A good fit was obtained for the
$E_{1g}$ case with an equal admixture of the two functions.  This result is
somewhat surprising since an equal admixture would imply (within a tight
binding framework) that the pair interaction has a substantial range (the
near neighbor separation is 7.8 a.u., but an (0,1,1) vector is 14.3 a.u.).
In the $E_{2u}$ case, we never found an adequate fit, although we do show a
typical result (equal admixture, but with opposite sign).  One should
be cautious, though, about ruling against an $E_{2u}$ model based on this,
since there are other functions involving non-primitive translation vectors,
with distances comparable to the (0,1,1) primitive vectors, which we have
not considered here (one at 13.4 a.u., another at 15.2 a.u.)

\section{Conclusions}

In conclusion, we have found by analyzing simple ellipsoidal models for
the Fermi surface that recent thermal conductivity data
cannot unambiguously differentiate between the $E_{1u}$ and $E_{2u}$
models for the symmetry of the order parameter.
Such a differentiation should be possible by looking at samples
with degraded quality, since in
the $E_{2u}$ case, the non-zero value of $\kappa_c(0)/\kappa_b(0)$ is
intrinsic, whereas for $E_{1g}$, it is due to impurities.  We have also
found that results using realistic Fermi surfaces and gaps with proper
lattice translational symmetry differ significantly from those based on
ellipsoidal Fermi surfaces, and have discovered an $E_{1g}$ tight binding gap
function which gives a good representation of the data.
We emphasize that the thermal conductivity
data appear to put great constraints on the overall shape of the gap
function, and thus will be an important
ingredient in determining the
validity of microscopic models for the superconductivity in $UPt_3$.

\acknowledgments

The authors gratefully acknowledge extensive
discussions with L. Taillefer and A. Fledderjohann, and in particular
the former for use of his group's unpublished data.  M.R.N.
was supported by the U.S. Dept. of Energy, Basic Energy Sciences, under
Contract No. W-31-109-ENG-38, and also thanks the University of Florida
for support in the initial stages of this work.

\begin{table}\caption{$\kappa_c(0)/\kappa_b(0)$ in the clean limit
for several $Y_{LM}$ gap functions for two values of the mass ratio, $m_r$.
The harmonics $Y_{21}$ and $Y_{41}$ have $E_{1g}$ symmetry and the harmonics
$Y_{32}$, $Y_{52}$, and $Y_{54}$ have $E_{2u}$ symmetry.}
\begin{tabular}{cccc}
LM & form & $m_r$=1 & $m_r$=2.8 \\
\tableline
10 & $\cos(\theta)$ & 0 & 0 \\
11 & $\sin(\theta)$ & $\infty$ & $\infty$ \\
20 & $3\cos^2(\theta)-r^2$ & 1 & 0.357 \\
21 & $\sin(\theta)\cos(\theta)$ & 0 & 0 \\
22 & $\sin^2(\theta)$ & $\infty$ & $\infty$ \\
30 & $\cos(\theta)(5\cos^2(\theta)-3r^2)$ & 6/7 & 0.423 \\
31 & $\sin(\theta)(5\cos^2(\theta)-r^2)$ & 1/2 & 0.180 \\
32 & $\sin^2(\theta)\cos(\theta)$ & 1 & 1 \\
33 & $\sin^3(\theta)$ & $\infty$ & $\infty$ \\
41 & $\sin(\theta)(7\cos^3(\theta)-3\cos(\theta)r^2)$ & 0.647 & 0.252 \\
52 & $\sin^2(\theta)(3\cos^3(\theta)-\cos(\theta)r^2)$ & 0.744 & 0.267 \\
54 & $\sin^4(\theta)\cos(\theta)$ & $\infty$ & $\infty$ \\
\end{tabular}
\end{table}

\begin{figure}
\caption{LDA Fermi surface for $UPt_3$ plotted in the symmetry planes of
the Brillouin zone.  The surface is composed of five bands, three centered
around $\Gamma$ and two centered around $A$.}
\label{fig1}
\end{figure}

\begin{figure}
\caption{$\kappa_c/\kappa_b$ (normalized to its value at $T_{c-}$) for various
harmonics through L=5 (curves labled by L,M) on an ellipsoid ($m_r$=2.8).
All plots, unless otherwise noted, are for an
impurity scattering rate, $\Gamma$, of 0.1 $T_c$ in the unitarity limit, with
an inelastic scattering rate $4T^2$ times the elastic rate.
The black dots are data from Ref. \protect\onlinecite{tai2}.}
\label{fig2}
\end{figure}

\begin{figure}
\caption{(a) $\kappa_c/\kappa_b$ and (b) $\kappa_i/T$ for mixed harmonics on
an ellipsoid ($m_r$=2.8).
The curve marked 1g is for $E_{1g}$ ($Y_{21}-0.15 Y_{41}$) and the one marked
2u for $E_{2u}$ ($Y_{32}+0.2 Y_{52}$).  Black dots in (a)
and black (white) dots in (b) are data
from Ref. \protect\onlinecite{tai2}.  The solid (dashed) curves in (b) are the
respective theoretical results for i=b (i=c).}
\label{fig3}
\end{figure}

\begin{figure}
\caption{Order parameter versus polar angle for mixed harmonics on
an ellipsoid ($m_r$=2.8).}
\label{fig4}
\end{figure}

\begin{figure}
\caption{$\kappa_c/\kappa_b$ at $T/T_c=0.02$ (solid line) and $T/T_c=0.2$
(dashed line) vs. normalized impurity scattering rate, $\Gamma/T_c$, for
pure ellipsoidal harmonics $Y_{32}$ (2u) and $Y_{21}$ (1g), as well as
for the mixed harmonics of Fig. 3a (2u$^*$ and 1g$^*$).}
\label{fig5}
\end{figure}

\begin{figure}
\caption{(a) $\kappa_c/\kappa_b$ and (b) $\kappa_i/T$ for Fermi surface
harmonics on the LDA Fermi surface.  Same notation as Fig. 3.}
\label{fig6}
\end{figure}

\begin{figure}
\caption{(a) $\kappa_c/\kappa_b$ and (b) $\kappa_i/T$ for tight binding
functions (0,1,1 type
primitive lattice vectors) on the LDA Fermi surface.  Same notation as Fig. 3.}
\label{fig7}
\end{figure}

\begin{figure}
\caption{(a) $\kappa_c/\kappa_b$ and (b) $\kappa_i/T$ for near neighbor tight
binding functions
(Konno-Ueda type) on the LDA Fermi surface.  Same notation as Fig. 3.}
\label{fig8}
\end{figure}

\begin{figure}
\caption{(a) $\kappa_c/\kappa_b$ and (b) $\kappa_i/T$ for mixed tight binding
functions on the LDA
Fermi surface.  Same notation as Fig. 3.}
\label{fig9}
\end{figure}


\begin{references}
\bibitem{ell} Some definitions of the harmonics are normalized by the factor
$r^L$, with $r$ no longer a constant in the ellipsoidal case.  Inclusion of
this
normalization increases the $E_{2u}$ value of $\kappa_c(0)/\kappa_b(0)$ by a
factor of over 3 for a mass ratio of 2.8.  We do not use this normalization
factor, since Allen's original paper on Fermi surface harmonics did not include
it (where $r$ would be replaced by the modulus of the Fermi velocity).  Still,
this ambiguity should be kept in mind.
\bibitem{pseudo} In the presence of spin-orbit, spin is no longer a good
quantum number, but one can define a pseudo-spin since the eigenvalues at
a general k point are doubly degenerate due to Kramers theorem.
\bibitem{fsh} Simply replacing $k_i$ by $v_i$ does not lead to orthogonal
functions, so higher L harmonics should be orthogonalized to lower L harmonics
of the same group representation.  For simplicity, we do not use these
more complicated functions here.
\bibitem{nlda} In this paper, results from a relativistic linearized muffin tin
orbital calculation (with combined corrections and basis functions
through l=3 on both U and Pt sites) were used.
\bibitem{tet2} First principle eigenvalues were calculated on a 721 k point
grid in the irreducible (1/24) of the Brillouin zone.  These were then fit
using a 2541 function Fourier spline series, which allows eigenvalues to be
determined at any k point, with velocities being constructed by simple
differencing.  The Fermi surface integrals were performed with a linear
tetrahedron method, with the zone broken down $72 \cdot 8^n$ times.  Most
calculations
used an n of 5, but some were done at an n of 6 to check for convergence.
\bibitem{fsh2} In general, these functions will be multiplied by a
constant (determined by solving the gap equations) which can vary from band
to band.  In this paper, we take this constant to be independent
of band index.
\bibitem{y54} $Y_{54}$ does not mix with $Y_{32}$ in axial symmetry.
\bibitem{hundx} The analysis in this paper was greatly facilitated by
the analytic nature of the basis functions.  Most microscopic models, such
as the one in Ref. \onlinecite{hund}, are numerical and thus more difficult
to treat given the large number of tetrahedra needed in the zone integrations.
\end{references}
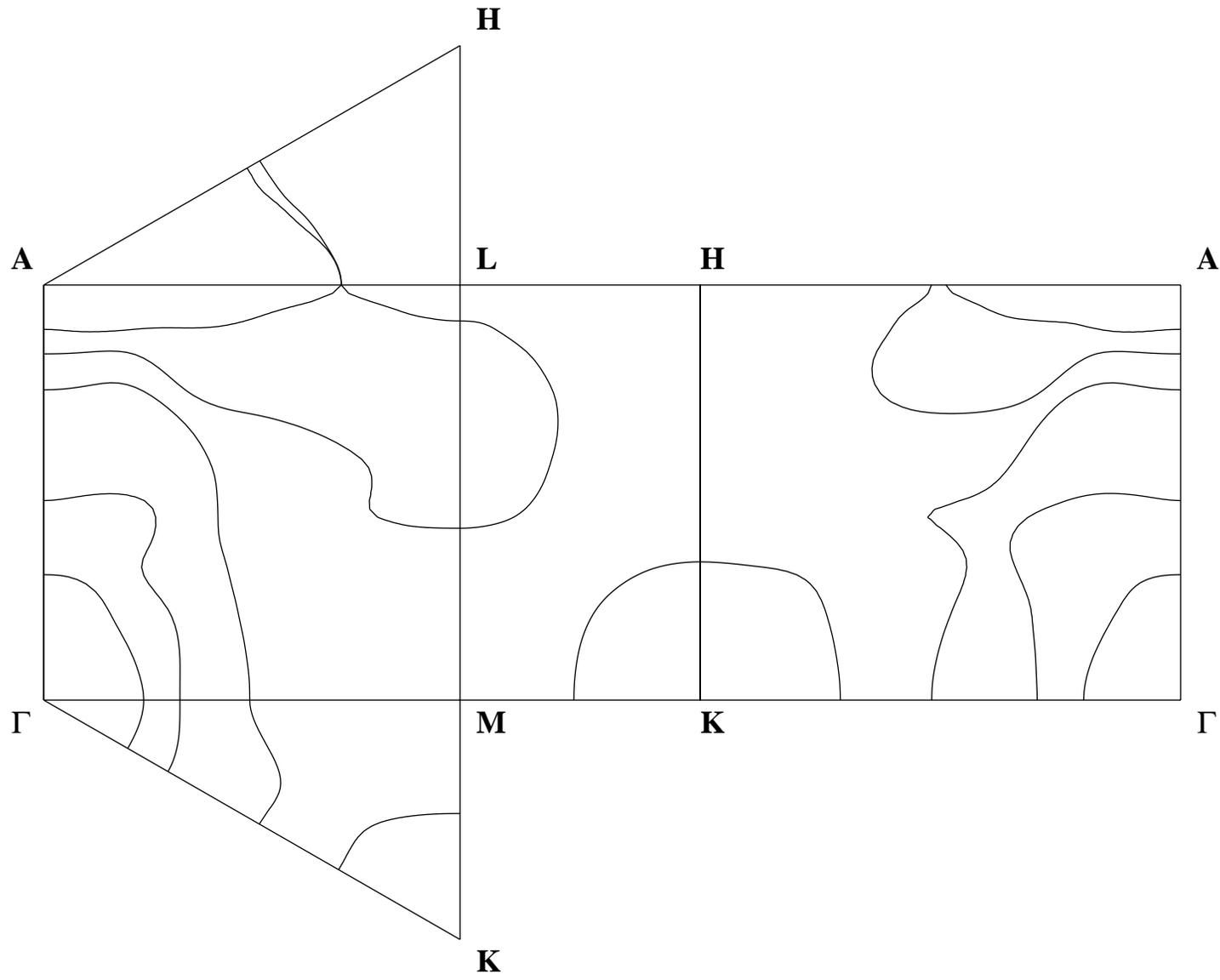
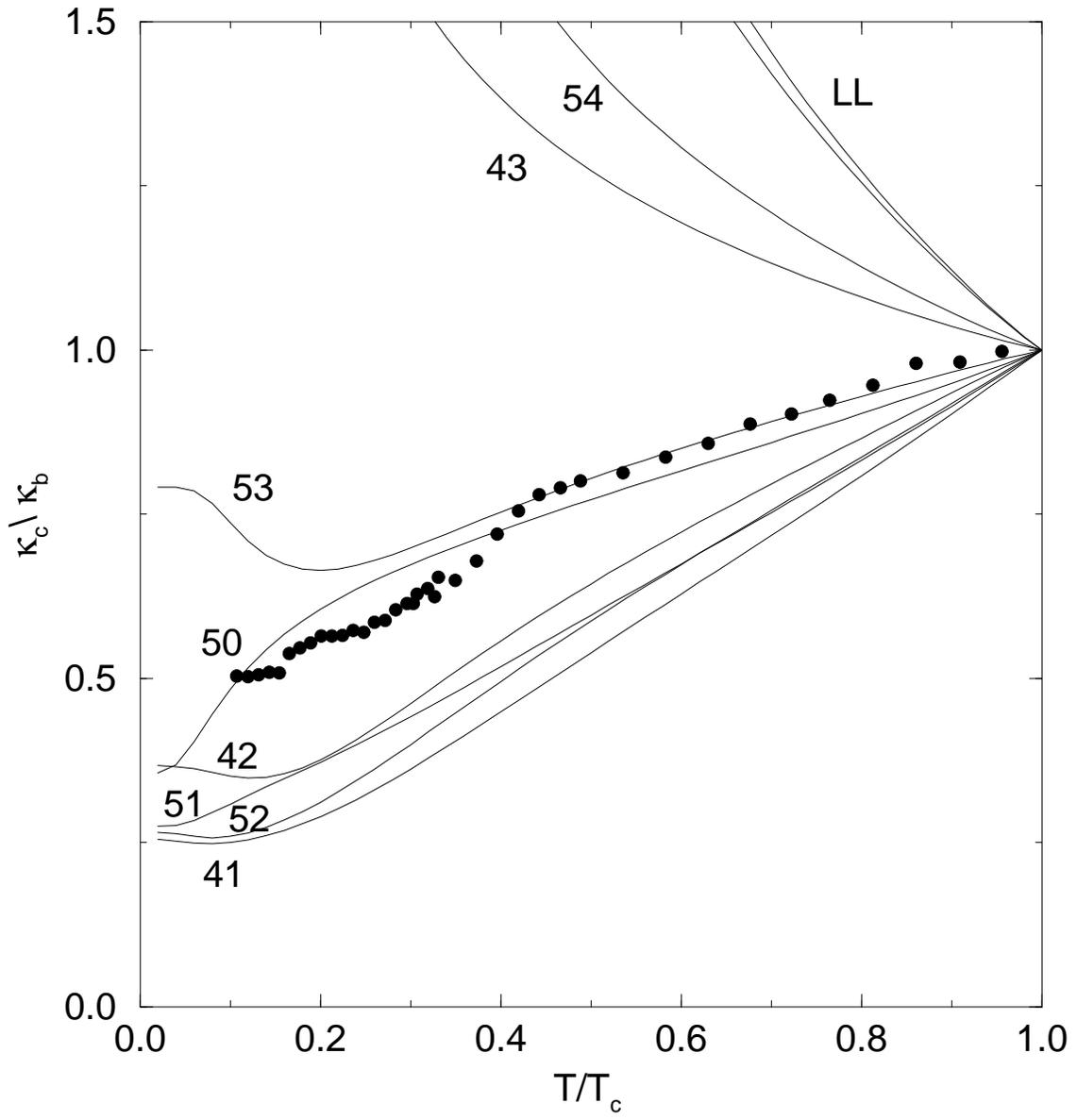
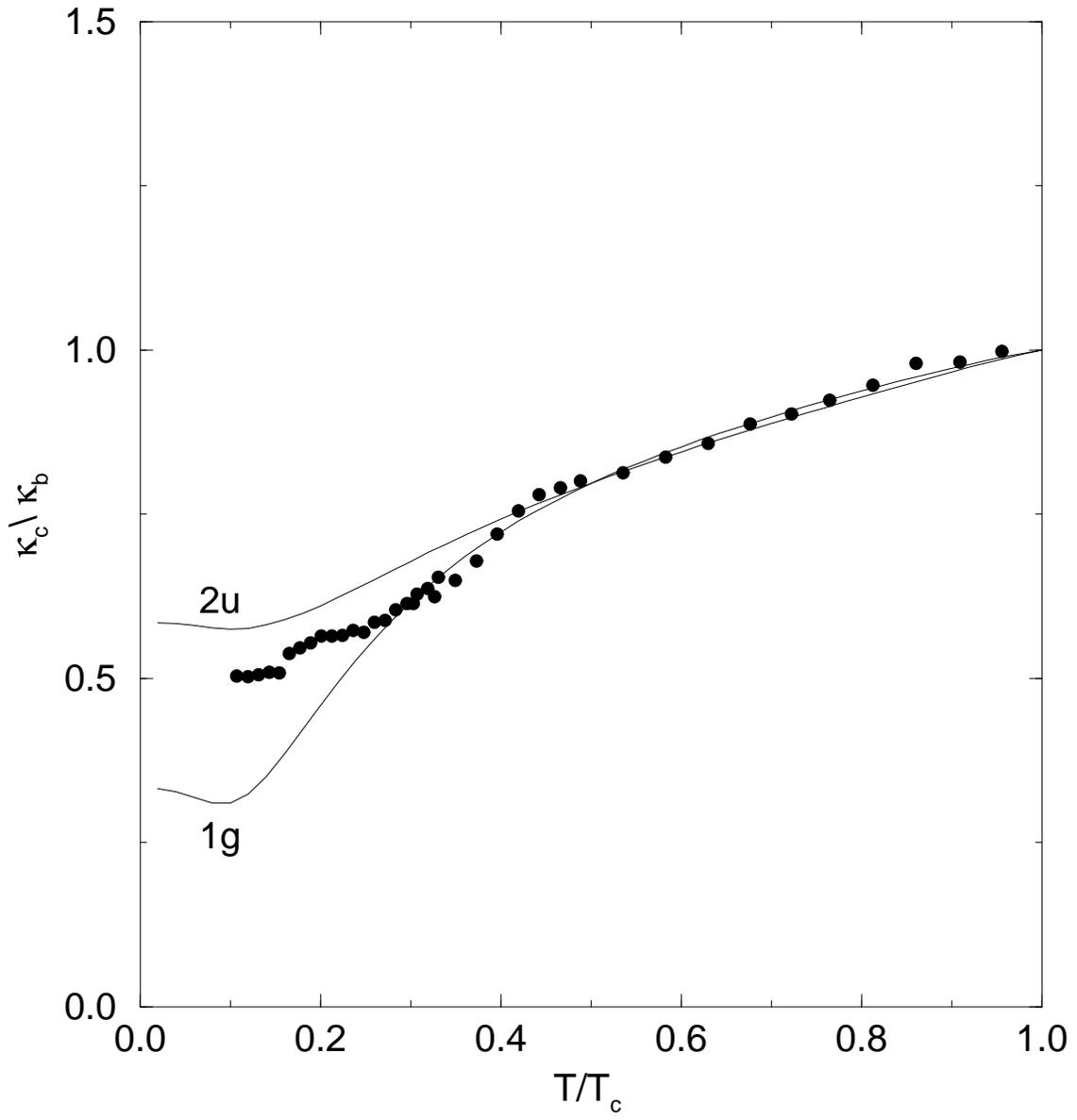
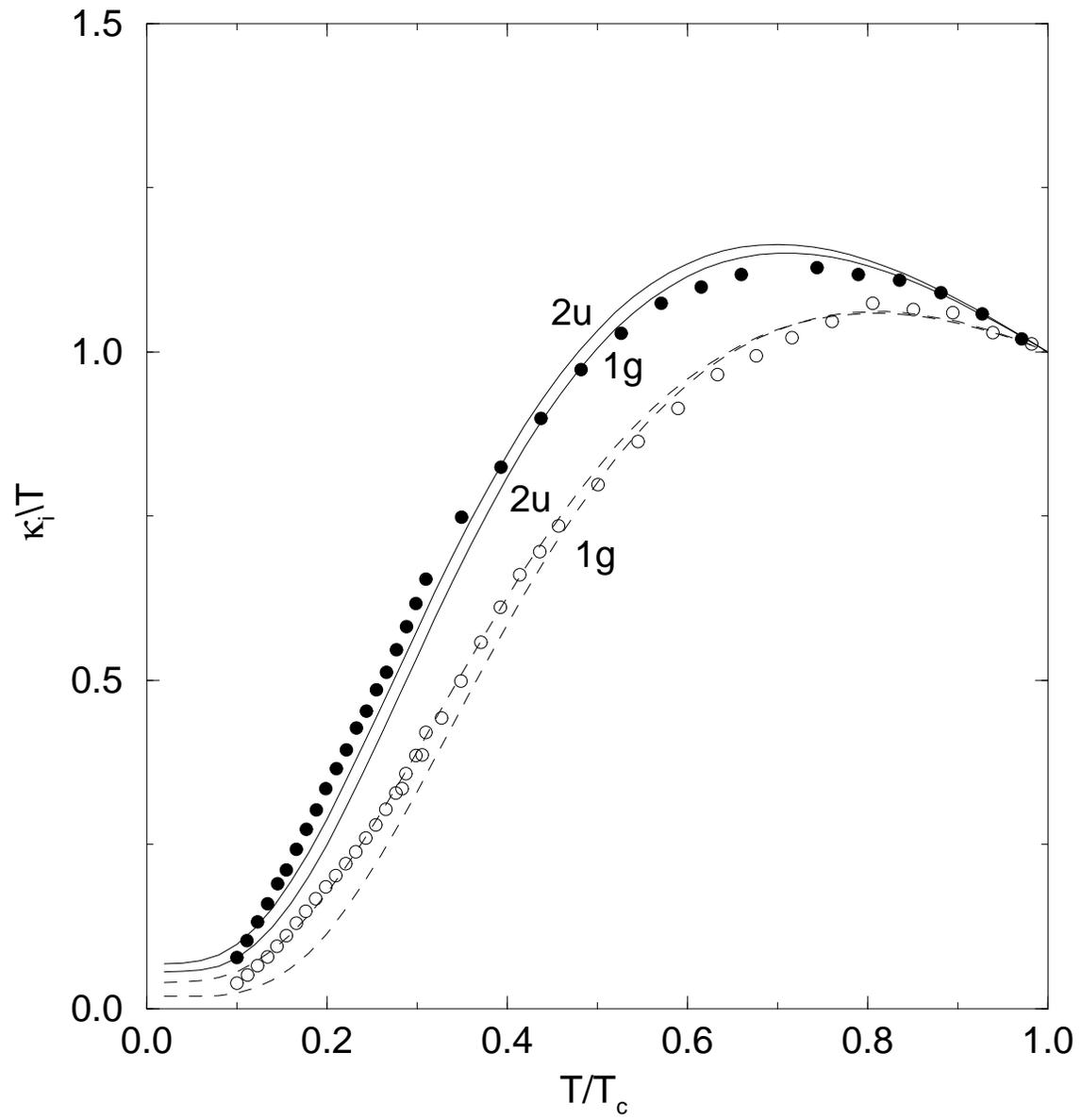
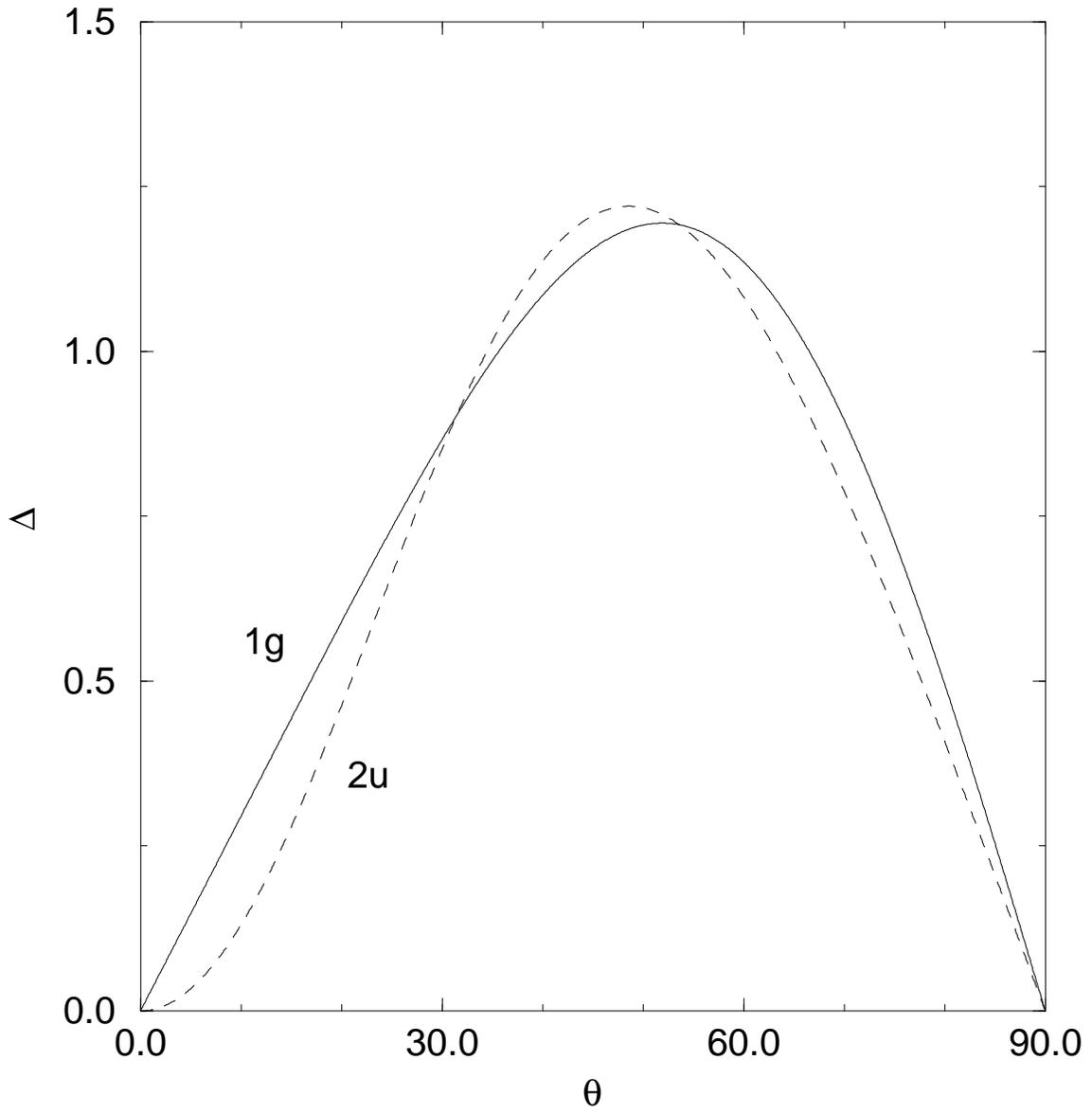
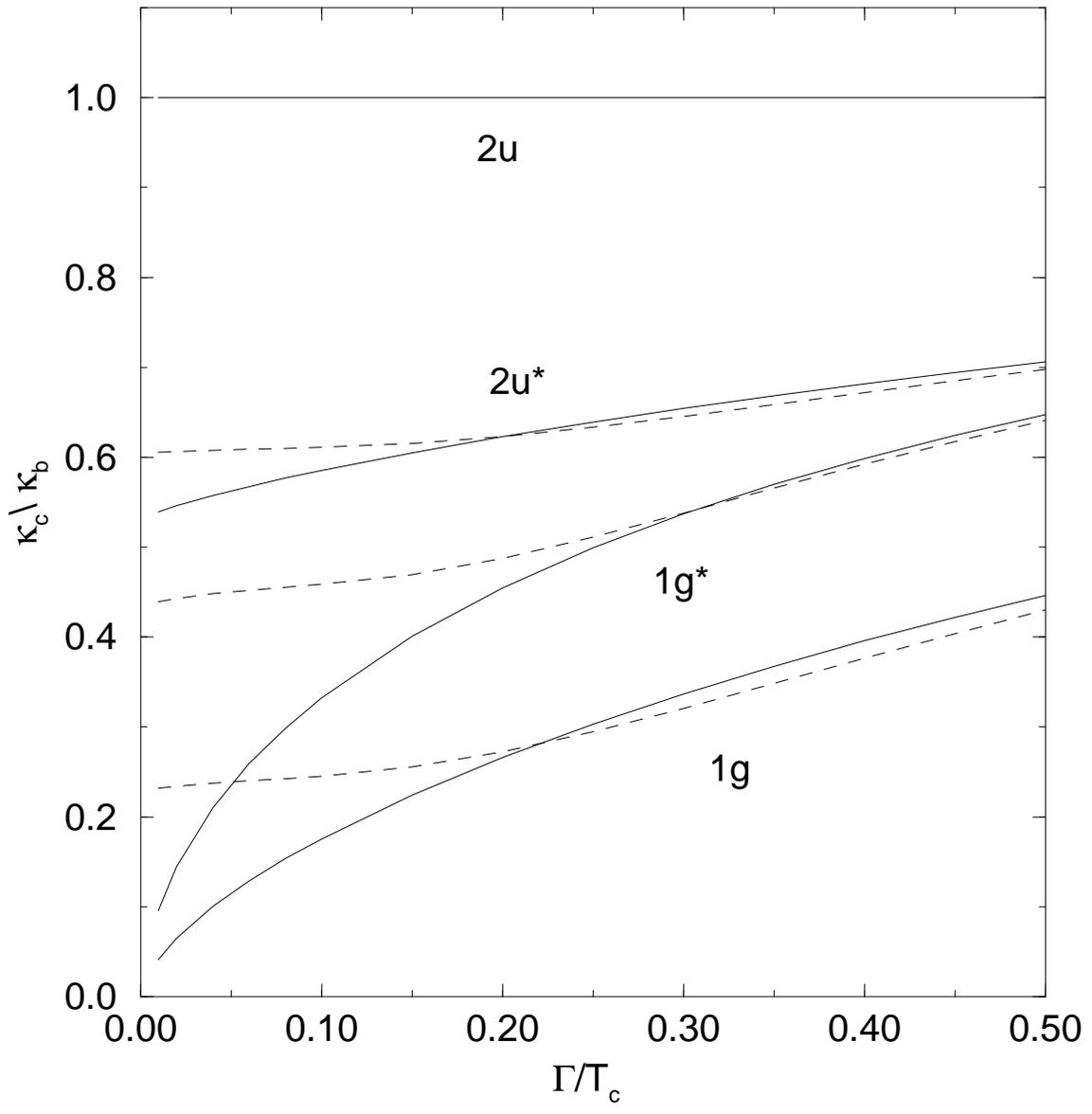
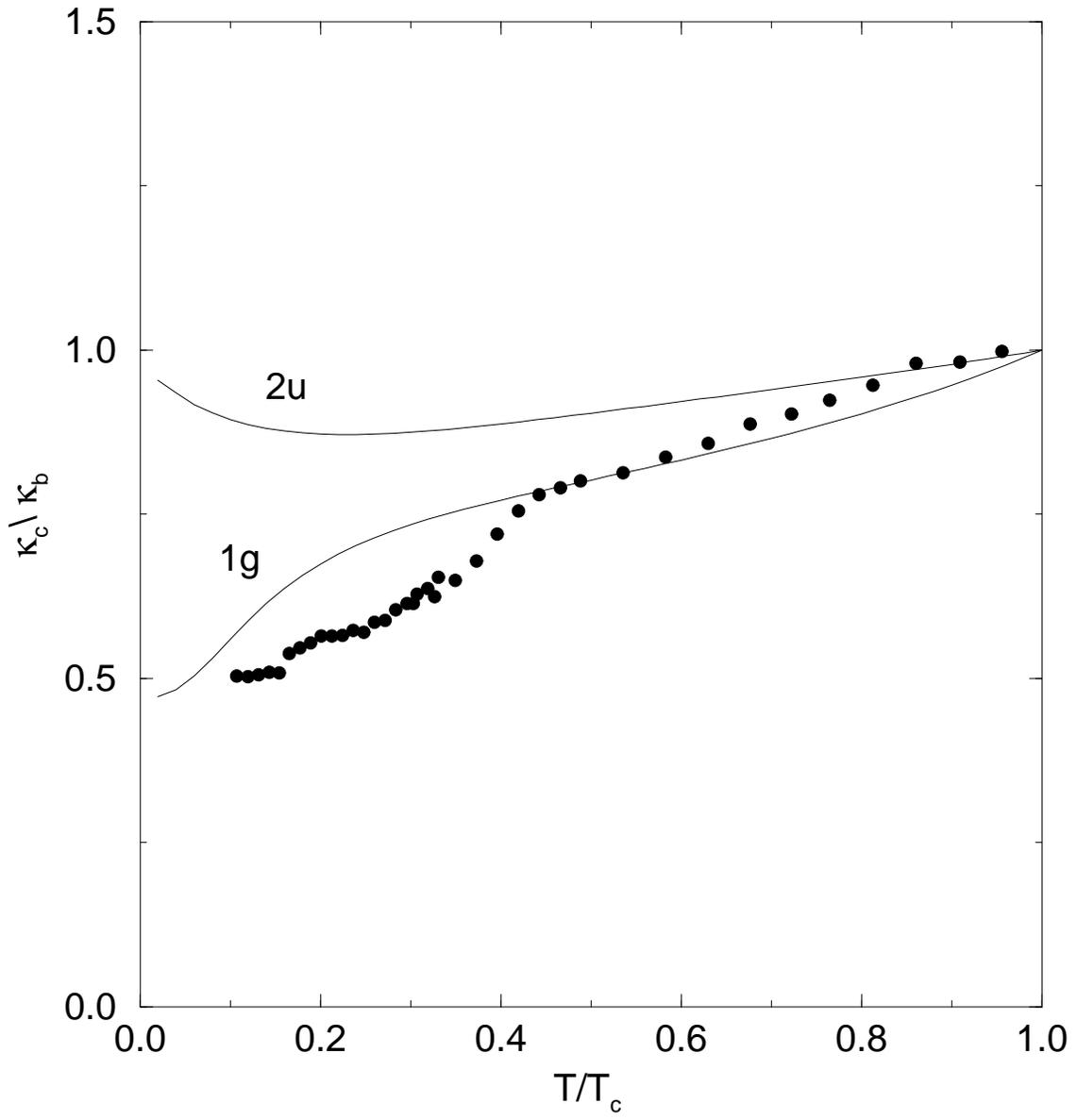
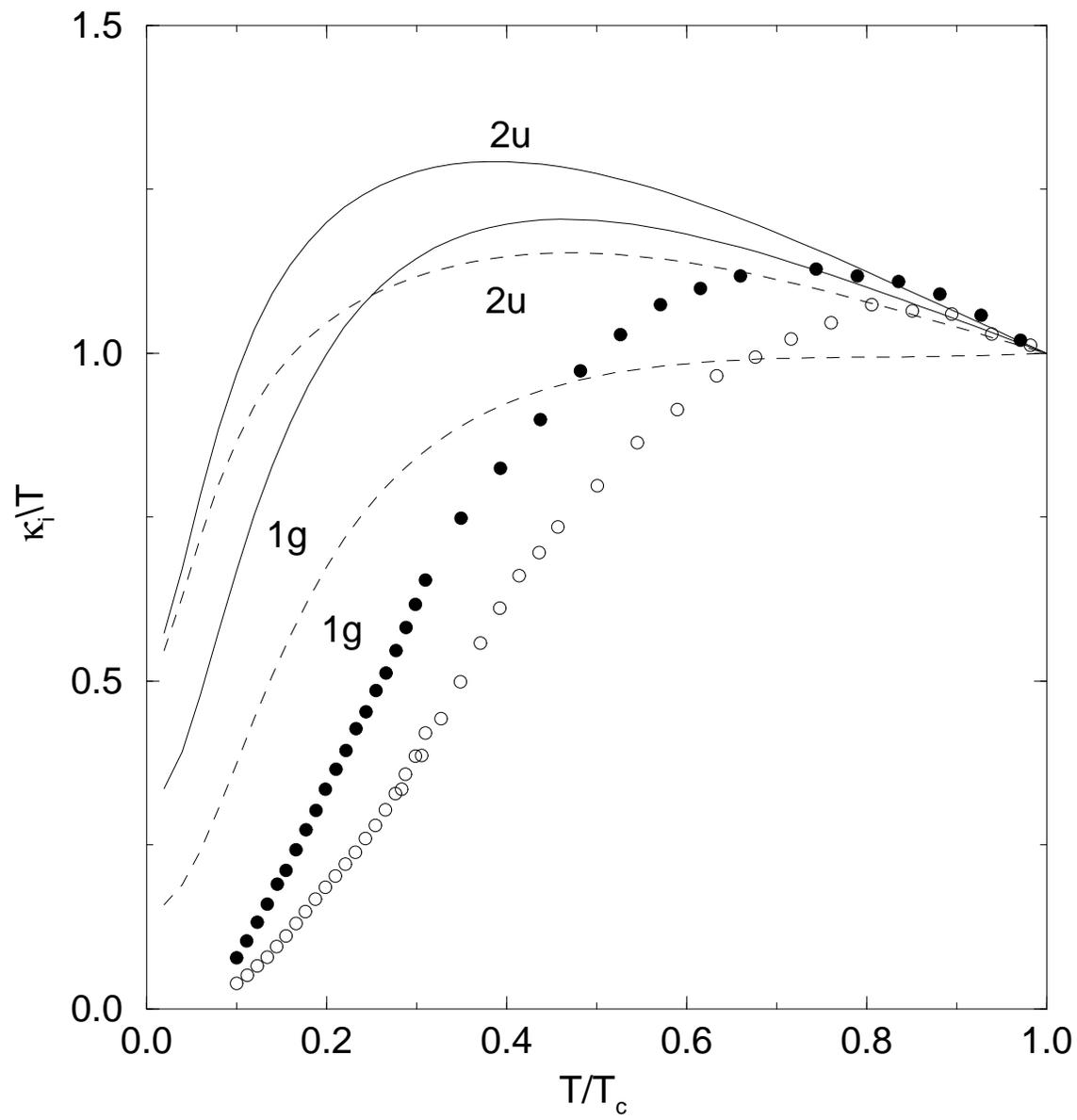
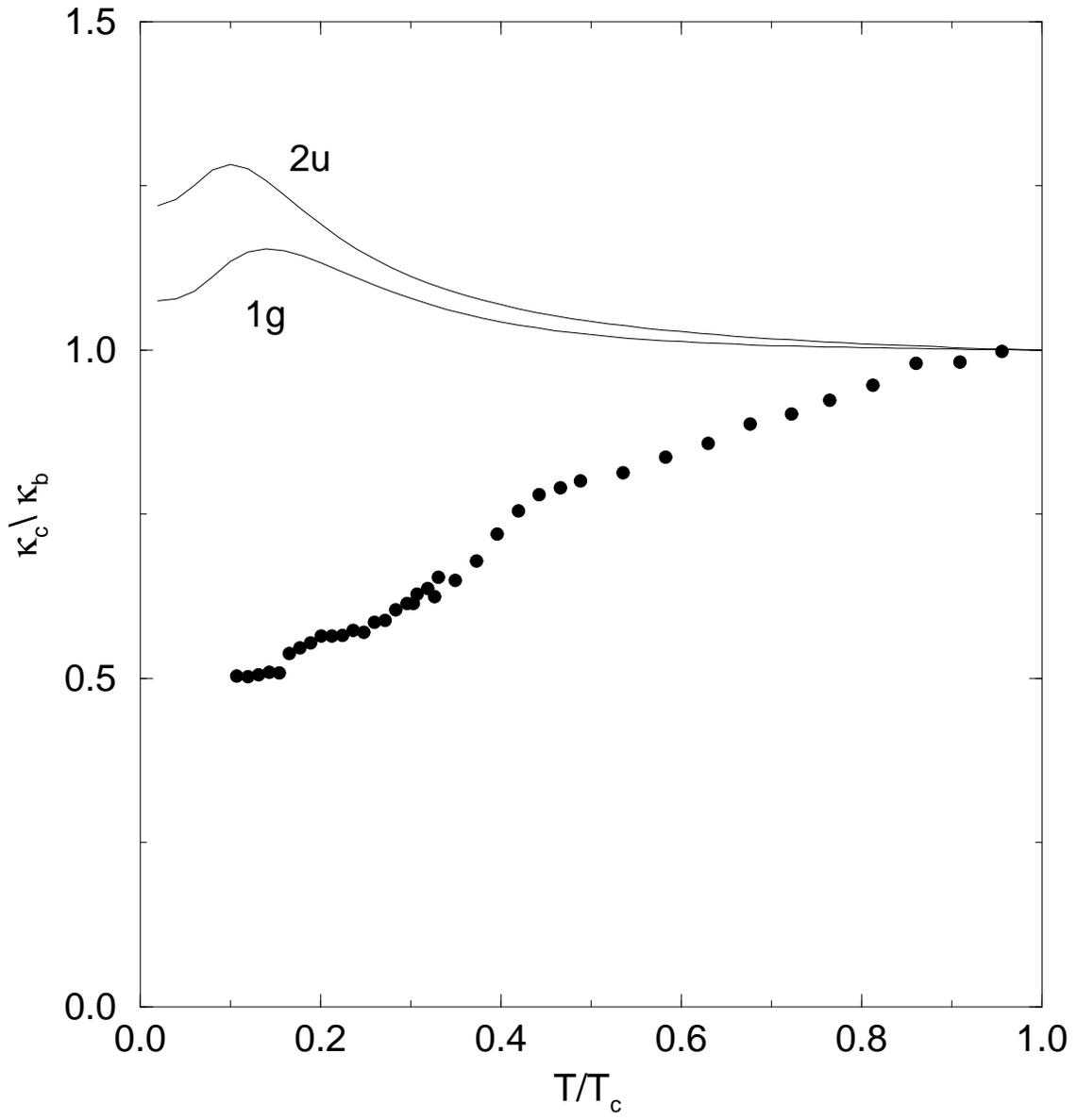
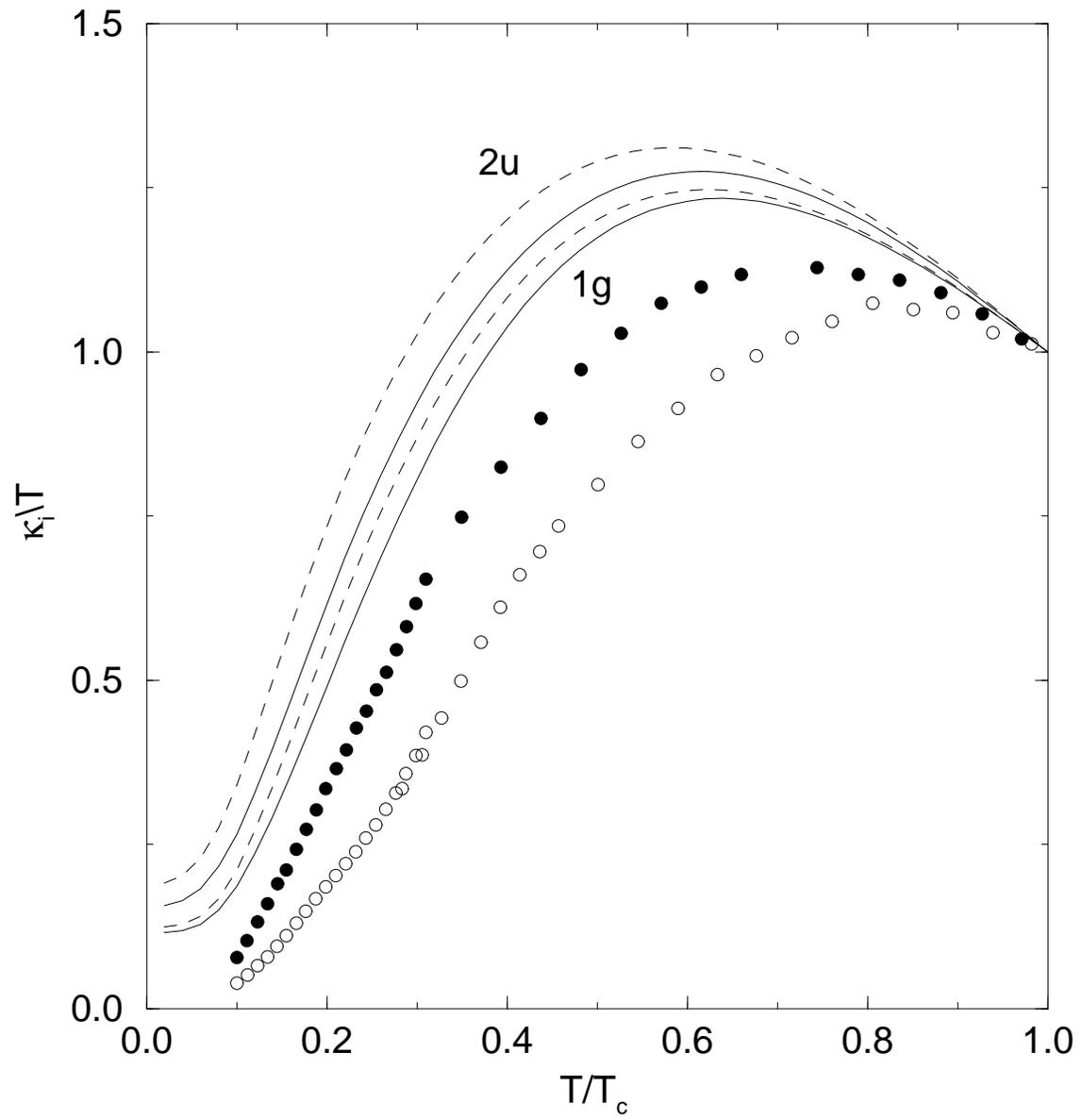
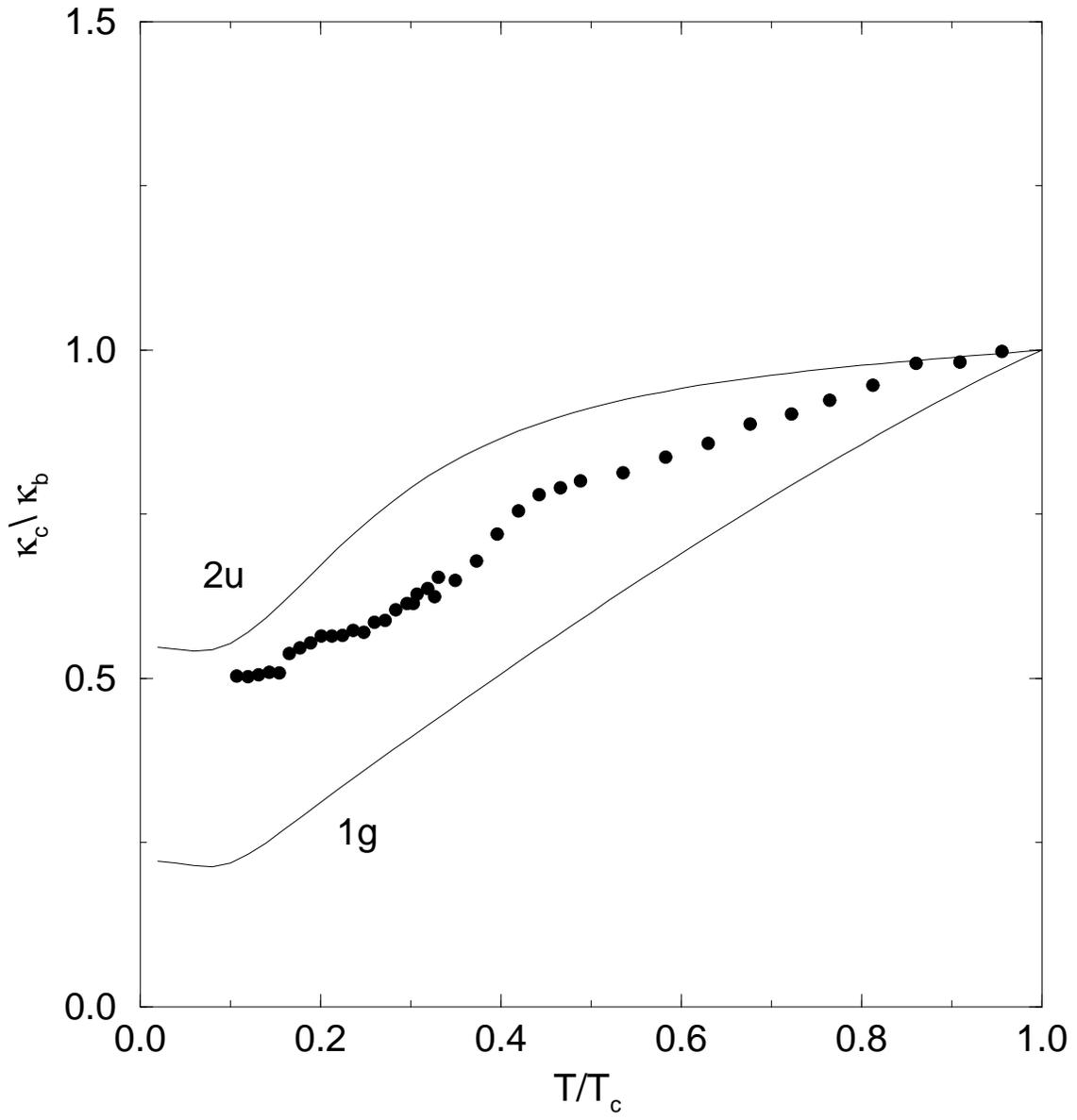
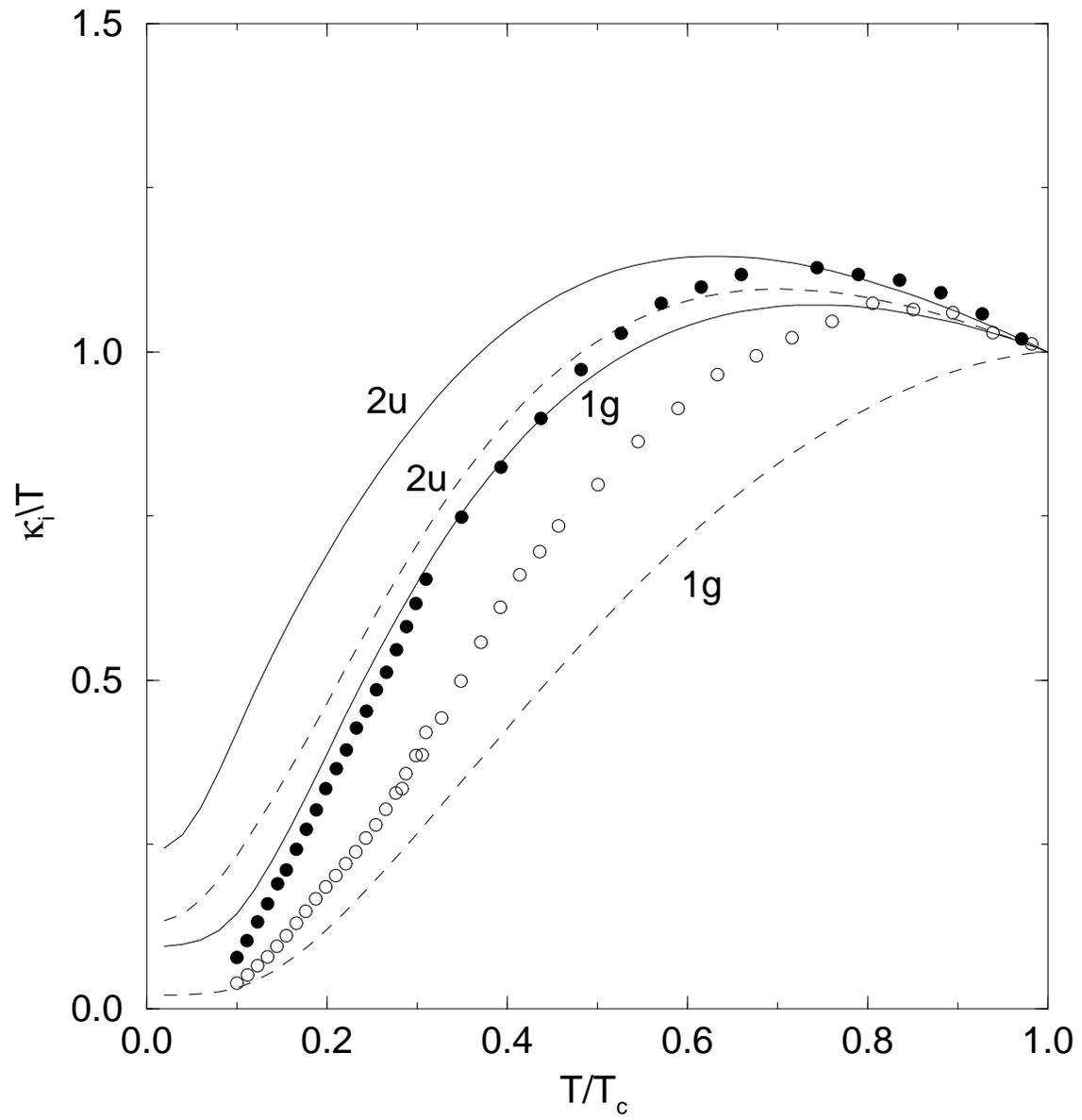
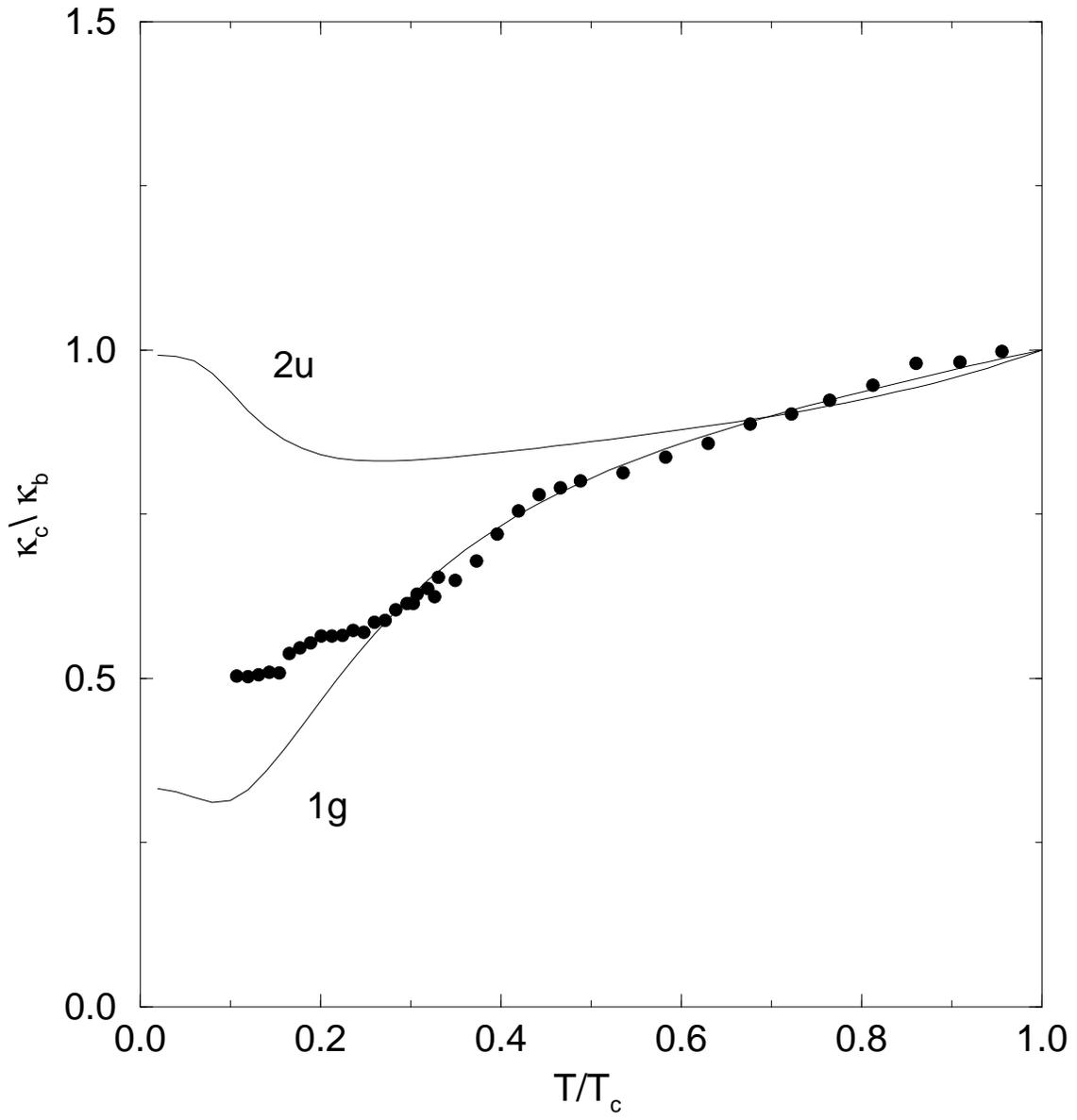
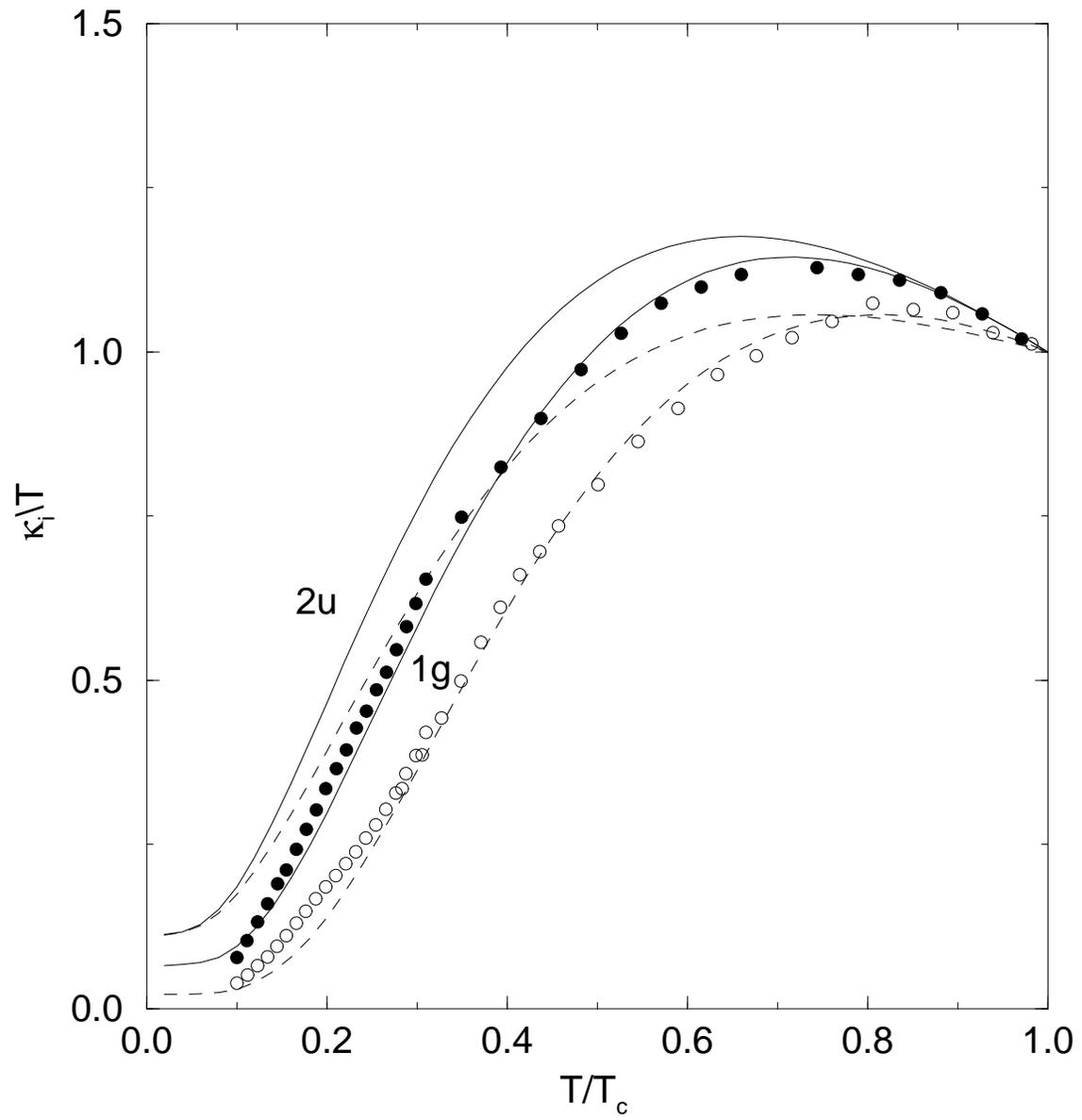
\end{document}